*A.V. Smirnov, T.V. Levashova, M.P. Pashkin, N.G. Shilov,*
*A.A. Krizhanovsky, A.M. Kashevnik, and A.S. Komarova*

# CONTEXT-SENSITIVE ACCESS
# TO E-DOCUMENT CORPUS[1]

## 1. Introduction

The methodology of context-sensitive access to e-documents considers context as a problem model based on the knowledge extracted from the application domain, and presented in the form of application ontology.

Efficient access to an information in the text form is needed. Wiki resources as a modern text format provides huge number of text in a semi formalized structure.

At the first stage of the methodology, documents are indexed against the ontology representing macro-situation. The indexing method uses a topic tree as a middle layer between documents and the application ontology. At the second stage documents relevant to the current situation (the abstract and operational contexts) are identified and sorted by degree of relevance. Abstract context is a problem-oriented ontology-based model. Operational context is an instantiation of the abstract context with data provided by the information sources.

The following parts of the methodology are described: (i) metrics for measuring similarity of e-documents to ontology, (ii) a document index storing results of indexing of e-documents against the ontology; (iii) a method for identification of relevant e-documents based on semantic similarity measures.

[1] This work was partly supported through project # 16.2.35 of the research program "Mathematical Modelling and Intelligent Systems", project # 1.9 of the research program "Fundamental Basics of Information Technologies and Computer Systems" of the Russian Academy of Sciences, the project funded by grant #05-01-00151 of the Russian Foundation for Basic Research. The presented research was partially supported through CRDF partner project # RUM2-1554-ST-05 with US ONR and US AFRL.

Wikipedia (wiki resource) is used as a corpus of e-documents for approach evaluation in a case study. Text categorization, the presence of metadata, and an existence of a lot of articles related to different topics characterize the corpus.

**2. Related work**

The ALVIS annotation format is designed for the indexing of topic-specific documents[1]. The linguistic annotation added to the textual entities (words, phrases, etc.) consists of: morpho-syntactic tags, syntactic relations, semantic tags (named entities, terms or undefined semantic units), and semantic relations (the anaphoric and domain specific relations). The segmentation of texts to tokens, words, phrases, semantic units (named entities, concepts specific to the domain, undefined units), and sentences is proposed. But at this moment the benefits of the ALVIS approach in comparison with, e.g., annotation and segmentation schemes proposed and implemented in GATE[2] are not evident.

The document classification problem was solved in the approach[3] by using the taxonomy of Wikipedia categories[4], since problem

---

[1] Nazarenko A., Alphonse E., Deriviere J., Hamon T., Vauvert G., Weissenbacher D. The ALVIS Format for Linguistically Annotated Documents // Proceedings of the fifth international conference on Language Resources and Evaluation, LREC 2006 (2006), pp. 1782-1786 **http://arxiv.org/abs/cs/0609136**.

[2] Cunningham H., Maynard D., Bontcheva K., Tablan V., Ursu C., Dimitrov M., Dowman M., Aswani N. and Roberts I. Developing language processing components with GATE (user's guide), Technical report, University of Sheffield, U.K., 2005. **http://www.gate.ac.uk**

[3] See topic-based index creation in Subsection 3.1.

[4] Voss J. Collaborative thesaurus tagging the wikipedia way // Collaborative Web Tagging Workshop, 2006.

orientation of a Wikipedia document is defined by a set of categories assigned to.

The document classification in the multilingual news analysis system NewsExplorer[1] is done via the Eurovoc thesaurus[2]. But NewsExplorer's developers had to perform preliminary manual classification of approximately 80 000 documents according to over 5 000 descriptors of Eurovoc.

Hence the main drawback of document classification based on Eurovoc thesaurus is initial manual tagging of documents in contrast to using of available tags (categories) assigned to Wikipedia articles. The Eurovoc's benefit is that the wide-coverage thesaurus descriptors (class names) exist in one-to-one translations in over twenty languages[3].

### 3. Approach

Document Index stores results of indexing the e-documents against the application ontology. One of the indexing purposes is to calculate and store the similarity of a document to a fragment[4] of the application ontology. The approach is described in our previous work[5] in detail.

---

**http://arxiv.org/abs/cs/0604036**

[1] Steinberger R., Pouliquen B., Ignat C. Navigating multilingual news collections using automatically extracted information // Proceedings of the 27th International Conference "Information Technology Interfaces" (ITI'2005), Cavtat / Dubrovnik, 2005. **http://arxiv.org/abs/cs/0609053**

[2] Eurovoc thesaurus, 2006 **http://europa.eu/eurovoc**

[3] Though Wikipedia exists (with categories) in 229 languages. 2006 **http://meta.wikimedia.org/wiki/List_of_Wikipedias**

[4] Ontology fragment contains sets of classes with relationships between them, class attributes, and domains for string attributes.

[5] Smirnov A., Levashova T., Shilov N., Pashkin M., Kashevnik A., Krizhanovsky A., Komarova A. Context-Sensitive Access to Information Sources // In Proceedings: International Conference on Hybrid Information Technology, Cheju Island, Korea, November 9th-11th, 2006. Accepted for publication.

The Index contains a set of tuples <$A^{II}$, *Doc_ID*, *Sim*>, where $A^{II}$ – application ontology fragment to which the document is similar; *Doc_ID* – document identifier; and *Sim* – value of similarity of the document to the application ontology fragment.

The initial set of ontology classes (for the document) is selected by comparing text similarity between class names and document's text / metadata. The problem is to select a set of ontology classes (as fragment of ontology) including all classes from the initial set.

*Ontology fragment selection algorithm* is proposed. It is given a set of classes $S_{CA}$ from application ontology[1,2]. The algorithm selects a subset of classes $M_{CL}$ connected by ontology relationships $M_{REL}$ (the attributes of classes $M_{CL}$ included in the fragment)[3]. This selection is based on the following rules: (1) $M_{CL}$ contains $S_{CA}$, (2) $M_{CL}$ and $M_{REL}$ contain the shortest path for each pair of classes in $S_{CA}$.

Thus, the algorithm takes set of classes and attributes (as an initial fragment) and searches the minimal subset of connected classes (adding classes and relations between them into the fragment).

### 3.1. Topic based index creation

The index is created between Wikipedia documents and ontology classes and attributes via Wikipedia categories. The algorithm has the following parameters:

WA – set of Wikipedia articles.

ONT – set of names of classes and attributes from ontology.

$D_{max}$ – maximum distance between similar words.

$O_{res}$ – result set of classes and attributes from ontology that are correspond to the document.

---

[1] The initial set of classes $S_{CA}$ is selected by comparing text similarity between class names and document's text.

[2] The text string similarity could be calculated by Levenshtein edit distance (number of insert, remove, replace character operations). See **http://en.wikipedia.org/wiki/Levenshtein_distance**).

[3] Thus, the fragment contains classes $M_{CL}$, their attributes, and ontology relationships $M_{REL}$.

k – weight coefficient in [0,1].
|CS$_{max}$| - number of categories in Wikipedia.

```
Input: Wikipedia_articles, ONT, D_max, O_res,
k, |CS_max|
Result: Index(a, O_res, sim)
begin
    Index ⟵ ∅
    for a ∈ Wikipedia_articles do
        O_res ⟵ ∅
1       D_pair ⟵ ∅
        d_sum = 0
2       CS ⟵ a ∪ Γ(a) ∪ Γ(Γ(a))
        for c ∈ CS do
            for o ∈ ONT do
                d = D_Levenshtein(c, o)
                if d < D_max then
                    O_res ⟵ o
                    D_pair ⟵ d
                    d_sum = d_sum + d
3       sim =  1 − ½ · [(1 − k) · |D_pair|/(|CS_max|·|ONT|)
                      + k · (1 − d_sum/(D_max·|D_pair|))]
4       Index ⟵ (a, O_res, sim)
end
```

*Fig. 1*. Topic-based indexing algorithm

Algorithm of topic-based index creation is designed and implemented. The short description of following steps is provided:
1) D $_{pair}$ is the array of Levenshtein distances between (1) names of ontology classes, attributes and (2) titles of articles, categories of Wikipedia.
2) The set of neighbours categories of categories of article *a* is added to *CS*.

3) The value of *sim* (step 11) is in the range [0,1], because:
- weight coefficient $k$ is in the range [0,1];
- $0 \leq |D_{pair}| / (|CS_{max}| \cdot |ONT|) \leq 1$ (size of subset of similar pairs). The size of $CS_{max}$ is a constant (number of categories in Wikipedia) here, but it is enough to require that $CS_{max}$ is not less than the maximum size of the set $|CS|$ for each article from WA in order to satisfy this inequality. But it will require some additional computations;
- $0 \leq d_{sum} / (D_{max} \cdot |D_{pair}|) \leq 1$ (ratio of Levenshtein distance sum to maximum possible distance from document text / metadata to ontology found element names $O_{res}$).
- let's sim:=1 if the array $D_{pair}$ (pairs of similar names of ontology classes and Wikipedia categories) is empty.
4) The set of found ontology classes and attributes (that have names similar to the title of Wikipedia article or categories of the article) is stored to Index.

The value of *sim* 1 means high similarity of set of ontology classes and attributes to the document, 0 – similarity is absent.

### *3.2. Relevance estimation*

At the second stage of approach, when Index $<A^{II}, Doc\_ID, Sim>$ was already created, the documents were evaluated by their relevance to the current situation. Measurement of similarity between contexts (as ontology model) and documents is based on a comparative assessment of these contexts and application ontology fragments $A^{II}$, elements of which have been included in the contexts.

Within the task of context-sensitive access to e-documents two methods for definition of documents relevant to abstract context have been proposed.

*In graph-based method* the fragment $A^{II}$ and abstract context are represented as two graphs. The graphs are compared taking into account (1) similarity between names of graph nodes; (2) number of neighbouring nodes for the similar nodes; (3) the shortest paths between similar nodes.

An algorithm implementing the graph-based method searches for similar nodes in the two graphs and compares the names of the nodes using Levenshtein distance. Nodes from the two graphs with the maximal degree of the similarity are connected into pairs. Thus, similar nodes in the two graphs are mapped as one-to-one. Then numbers of the neighbouring nodes for each pair are compared. The greater the difference between the numbers of the neighbouring nodes for each node in a pair the less the similarity for this pair. The path between pairs of similar nodes in each graph are calculated and then paths lengths are compared. The greater the difference between the paths lengths the less similarity of the nodes in the pair. The algorithm returns a numerical value in range [0, 1] of similarity between the two graphs.

The fragments are sorted according to their similarity to the abstract context. The fragments having greater similarity are considered as more relevant in this particular situation. Since for each document there is a fragment corresponding to it, documents are sorted by their relevance to the abstract context as well.

A *weight-based method* is developed for application ontology (AO) fragment specified as <$e_j$, $w_{ji}$, $kd_{ji}$>, where $e_j$ – an element of AO (class $o_j$ or attribute $q_j$) for which (1) the element name, or (2) synonym of the element name, or (3) the part of the element description (excluding stop words) is contained in the document content, $w_{ji}$ – weight of element $e_j$ within document $i$ for the given set of documents, and $kd_{ji}$ – coefficient of similarity of the description given for element $e_j$ in document $i$.

For the abstract context weights ($wa_j$) of AO element $e_j$ are calculated. The weight of the classes included in the context is 1. The set of the classes included in the context is extended with the classes from AO, chosen based on the shortest path from the classes to the abstract context. To find the shortest path from the classes to the abstract context the associative, taxonomical and hierarchical relationships of AO are taken into account. The weight of the class is inversely proportional to the appropriate shortest path, and the value of the shortest path is restricted by the expert. Weights of the attributes of AO are equal to the weights of the classes they belong to.

The similarity ($s$) between the document and the context is based on the data stored in the Index and on AO element weights ($wa_j$). The similarity is calculated using the angular separation metric. This metric is based on two vectors: the first one consists of AO element weights ($wa_j$) and the second vector consists of element weights ($w_{ji}$) from the Index with taking into account coefficient of similarity ($kd_{ji}$) from the Index between the description given for element and document.

As the result the relevance of the document to the context is calculated. The relevance is based on two variables: (1) the similarity ($s$) between the document and the context, (2) the coefficient of similarity ($Sim$) between the document metadata and the ontology from the Index.

### 5. Conclusion

Context-sensitive access to e-documents approach is described in the paper. The ontology fragment selection algorithm is presented briefly. The structure of index (documents are indexed against the ontology) and topic-based indexing algorithm are described. Two methods for identification of relevant e-documents based on semantic similarity measures are described: graph-based method (with graph comparison algorithm) and weight-based method.

The prototype has the following disadvantage. The linguistics annotation of texts is not used now, since the current prototype uses some predefined templates for user request defining. The user request in free form will require adding of linguistics annotation at the step of indexing the documents.

The logical evolution of the current approach is a clustering of similar documents for search improving. It is possible, since similarities of documents against fragments of problem-oriented ontologies are calculated. Thus, a function aggregating these similarities as a basis for documents clustering should be developed.

The document classification could be improved by using the linguistics resources such as, Eurovoc and WordNet. At the same time, the possibility to cluster documents should allow classifying new documents, since (1) Wikipedia document classified via Wikipedia categories, (2) Wikipedia documents and new (non Wikipedia) documents will share the same cluster, due to huge number of Wikipedia articles in different languages for different problem-oriented domains.